%% file: Index.tex
\documentclass[preprint,12pt]{elsarticle}
\usepackage{graphicx}
\usepackage{amssymb}
\usepackage{lineno}
\usepackage{bm}
\usepackage{color}

\usepackage[utf8]{inputenc}
\usepackage[T1]{fontenc}
\usepackage{mathptmx}
\usepackage{amsmath,amssymb}
\usepackage{etoolbox}
\usepackage{braket}
\usepackage{ulem}
\usepackage{url}

\makeatletter
\def\@email#1#2{%
 \endgroup
 \patchcmd{\titleblock@produce}
  {\frontmatter@RRAPformat}
  {\frontmatter@RRAPformat{\produce@RRAP{*#1\href{mailto:#2}{#2}}}\frontmatter@RRAPformat}
  {}{}
}%
\makeatother

\usepackage{multirow}
\usepackage{array}
\newcolumntype{C}[1]{>{\hfil}m{#1}<{\hfil}}
\makeatletter

    \@addtoreset{equation}{section}
\makeatother

\begin{document}

\input{Maintext}

\clearpage
\input{Supplementary}

\end{document}

%% file: Maintext.tex
%\documentclass[preprint,12pt]{elsarticle}
%\usepackage{graphicx}
%\usepackage{amssymb}
%\usepackage{lineno}
%\usepackage{bm}
%\usepackage{color}
%\newcommand{\bbm}[1]{{\mbox{\boldmath $#1$}}}
%\usepackage[utf8]{inputenc}
%\usepackage[T1]{fontenc}
%\usepackage{mathptmx}
%\usepackage{amsmath,amssymb}
%\usepackage{braket}
%\usepackage{ulem}
%\usepackage{url}
%\newcommand{\average}[1]%{\ensuremath{\langle#1\rangle} }

%\journal{Computer Physics Communications}

%\begin{document}

\begin{frontmatter}

\title{Thermal Conductivity Calculation using Homogeneous Non-equilibrium Molecular Dynamics Simulation 
with Allegro}

\author[1]{Kohei Shimamura\corref{cor1}}
\ead{shimamura@kumamoto-u.ac.jp.}
\cortext[cor1]{Author to whom correspondence should be addressed}
\address[1]{Department of Physics, Kumamoto University}

\address[2]{Advanced Research Laboratory, Technology Infrastructure Center, Technology Platform, Sony Group Corporation}

\address[3]{Collaboratory for Advanced Computing and Simulations, University of Southern California}

\author[2]{Shinnosuke Hattori}

\author[3]{Ken-ichi Nomura}

\author[1]{Akihide Koura}

\author[1]{Fuyuki Shimojo}

\begin{abstract}
In this study, we derive the heat flux formula for the Allegro model, one of machine-learning interatomic potentials using the equivariant deep neural network, to calculate lattice thermal conductivity using the homogeneous non-equilibrium molecular dynamics (HNEMD) method based on the Green-Kubo formula.
Allegro can construct more advanced atomic descriptors than conventional ones, and can be applied to multicomponent and large-scale systems, providing a significant advantage in estimating the thermal conductivity of anharmonic materials, such as thermoelectric materials.
In addition, the spectral heat current (SHC) method, recently developed for the HNEMD framework (HNEMD-SHC), allows the calculation of not only the total thermal conductivity but also its frequency components.
The verification of the heat flux and the demonstration of HNEMD-SHC method are performed for the extremely anharmonic low-temperature phase of Ag$_2$Se.
\end{abstract}


\end{frontmatter}

%% main text
\section{\label{sec:intro}Introduction}
The computational framework based on the Green-Kubo (GK) formula for molecular dynamics (MD) simulations using machine-learning interatomic potential (MLIP), has attracted attention in estimating the lattice thermal conductivities of various materials~\cite{Liang_2023,Brorsson_2022,Tisi_2021,Fan_2021,Verdi_2021,Han_2021,Li_2020,Huang_2019,Korotaev_2019,Shimamura_2021}.
The GK formula is as follows:
\begin{eqnarray}
\kappa &=& \frac{\varOmega}{3k_{\rm B}T^2}\int^{\infty}_{0} \langle{\bf{J}}_Q(t) \cdot {\bf{J}}_Q(0)\rangle dt
\label{eq:kappa},
\end{eqnarray}
where $k_{\rm B}$, $T$, $\varOmega$, and ${\bf{J}}_Q$ are the Boltzmann constant, temperature, volume of the supercell, and heat flux, respectively.
Two significant advantages of MLIP underpin this framework: low computational cost and high accuracy achieved through training on first-principles MD (FPMD) data.
While this framework is employed to explore and investigate materials with low thermal conductivity such as thermoelectric materials, MLIPs that can precisely describe complex atomic motions with strong anharmonicity are required to estimate thermal conductivity more accurately. 
In particular, appropriate descriptors that can characterize the atomic motions should be designed.

The recently proposed Allegro model~\cite{Musaelian_2023} is a useful MLIP candidate for thermal conductivity calculations.
The descriptors used in Allegro, similar to many conventional descriptors such as the Behler-Parrinello symmetry functions~\cite{Behler_2007}, characterize the local atomic environment, however, with significant improvements.
Allegro is based on atomic cluster expansion (ACE)~\cite{Drautz_2019}, which allows the construction of descriptors that incorporate body-order (BO) expansion.
Although most conventional descriptors consist of up to three-body orders (e.g., radial and angular parts) constrained by computational costs, four or more body orders can be incorporated.
Allegro also can efficiently generate descriptors with higher-order rotational symmetry with the aid of rotational equivariance.
Furthermore, although conventional descriptors require careful adjustment of the number of descriptors and functional forms of several hyperparameters, in Allegro the order of BO expansion and rotational symmetry can be controlled with $N_{\rm layer}$ and $l_{\rm max}$ parameters, respectively.
Note that Allegro is constructed using a graph neural network (GNN) with $N_{\rm layer}$ corresponding to the number of convolutional layers. 
The use of GNNs also solves the cumbersome problem of exponential increase in the number of conventional descriptors with an increase in the number of atomic species, leading to a high computational cost.
This is an indispensable capability for studying thermoelectric candidates in multicomponent systems.

However, Allegro does not implement the appropriate atomic virial tensor for the heat flux ${\bf{J}}_Q$:
\begin{eqnarray}
{\bf{J}}_Q &=& \frac{1}{\varOmega}\sum^{N_{\rm atom}}_{i} t_i{\bf{v}}_i \
                + \frac{1}{\varOmega}\sum^{N_{\rm atom}}_{i} \epsilon_i{\bf{v}}_i \
                + \frac{1}{\varOmega}\sum^{N_{\rm atom}}_{i} {\rm W}_i{\bf{v}}_i 
\label{eq:hflux},
\end{eqnarray}
where $t_i$, ${\bf{v}}_i$, $\epsilon_i$, and ${\rm W}_i$ are the atomic kinetic energy, velocity, potential energy, and virial tensor of the $i$th atom, respectively.
$N_{\rm atom}$ denotes the number of atoms in the system.
The atomic virial tensor plays an essential role, because its contribution accounts for the majority of the thermal conductivity~\cite{Shimamura_2021,Takeshita_2022,Andrade_2004}.
In addition, recent studies~\cite{Fan_2015,Surblys_2019, Boone_2019} report that the atomic virial tensor can have a strong effect on the accuracy of the thermal conductivity especially when dealing with many-body potentials, including MLIPs.
While an infinite number of atomic virial tensors can be defined in the many-body potential, an appropriate tensor must be used for the heat flux~\cite{Shimamura_2021,Shimamura_2020}.
Therefore, in this study, we derive the atomic virial tensor appropriate for the heat flux of Allegro and attempt to calculate the thermal conductivity based on the GK formula in Allegro.
We used the extended version of Allegro proposed by Yu et al.~\cite{Yu_2023}.
The test system was the highly anharmonic, low-temperature phase of Ag$_2$Se ($\beta$-Ag$_2$Se)~\cite{Wiegers_1971}, and we investigated the dependence of $l_{\rm max}$ and $N_{\rm layer}$ on the atomic structure and thermal conductivity.

The homogeneous non-equilibrium MD (HNEMD) method~\cite{Ciccotti_1979,Evans_1982,Yoshiya_2004} based on the GK formula was employed.
This method has the advantage of saving computational time compared to calculating the GK formula directly and can decompose the thermal conductivity into its partial contributions~\cite{Fujii_2020,Fan_2019}.
In particular, the recently proposed scheme combined with the spectral heat current (SHC) method (hereafter referred to as HNEMD-SHC)~\cite{Fan_2019} allows the analysis of thermal conductivity in frequency space, even for highly anharmonic materials.
We used the HNEMD-SHC method along with the vibrational density of states (VDOS)~\cite{Loong_1992}, for the thermal conductivity analysis.

In Section~\ref{sec:methods}, we explain the computational methods, such as the creation of training data for $\beta$-Ag$_2$Se, construction of Allegro models with $N_{\rm layer}$ and $l_{\rm max}$ parameters, formulas of the atomic virial tensors for Allegro, thermal conductivity calculation using the HNEMD method, and its decomposition in frequency space using the HNEMD-SHC method.
In Section~\ref{sec:results}, we first discuss the accuracy of the Allegro models using not only root mean square errors in training, but also physical quantities obtained from the MD simulations, such as stress and radial distribution functions.  
Second, we discuss the accuracy of the total thermal conductivities, comparing them with experimental value of $\beta$-Ag$_2$Se, using the HNEMD-SHC method to explain the differences in total thermal conductivity between the Allegro models in frequency space.
Finally, conclusions are presented in Section~\ref{sec:conclusion}.

\section{\label{sec:methods} Computational Methods}
\subsection{\label{ssec:bag2se}Generation of $\beta$-Ag$_2$Se training data by FPMD simulation }
All the FPMD and MD simulations using the Allegro models in this study were executed using the QXMD code~\cite{Shimojo_2019}.
Common to all the MD simulations is the time step $\Delta t$ that was set to 2.42 fs, performing at 300 K and 0 GPa.
In addition, a system comprising 256 Ag and 128 Se atoms was used. 

In the FPMD simulations, the electronic states were calculated using the projector augmented wave method~\cite{Blochl_1994,Kresse_1999} within the framework of the density functional theory (DFT)~\cite{Hohenberg_1964,Kohn_1965}.
Projector functions were generated for the 4$d$, 5$s$, and 5$p$ states of Ag and for the 4$s$, 4$p$, and 4$d$ states of Se.
The Perdew-Burke-Ernzerhof generalized gradient approximation~\cite{Perdew_1996} was used for the exchange correlation energy.
To correctly represent the electronic states in the localized $d$ orbitals of Ag, the DFT+U method~\cite{Anisimov_1997} with an effective parameter for the Coulomb interaction $U_{\rm eff} = 6.0$ eV~\cite{Fukushima_2019_2, Santamaria_2012} was used.
An empirical correction of the van der Waals interactions using the DFT-D approach~\cite{Grimme_2006} was employed.
The plane-wave cutoff energies were 20 and 200 Ry for the electronic pseudo-wave function and pseudo-charge density, respectively.
The energy functional was minimized iteratively using a preconditioned conjugate-gradient method~\cite{Shimojo_2001}.
The $\Gamma$ point was used for Brillouin zone sampling.

The training data were generated for $\beta$-Ag$_2$Se at 300 K and 0 GPa by performing FPMD simulations.
The unit cell of $\beta$-Ag$_2$Se has an orthorhombic structure consisting of four Ag$_2$Se groups with lattice parameters: $a = 4.333$ \AA, $b = 7.062$ \AA, and $c = 7.764$ \AA~\cite{Wiegers_1971}.
32 unit cells were arranged to construct an atomic configuration similar to cubic structure containing 256 Ag and 128 Se atoms, as shown in Fig. S1 in Supplementary Material (SM).
The lattice parameters of supercell were $L_1 = 17.332$ \AA, $L_2 = 20.991$ \AA, $L_3 = 20.991$ \AA, $\alpha = 84.58^\circ$, $\beta = 90.00^\circ$, and $\gamma = 90.00^\circ$.
Using the configuration under periodic boundary conditions, an FPMD simulation with an isothermal and isobaric ($NPT$) ensemble 
was performed in 4,100 steps.
A total of 820 MD steps were used as training data by extracting every five steps.
The $I$th MD step data contained the atomic coordinates \{${\bf{r}}_{I,i}$\}, 
total potential energy $E_I$, atomic force \{${\bf{F}}_{I,i}$\}, virial tensor ${\rm W}_{I}$, and supercell volume $\varOmega_{I}$.

\subsection{Allegro model}
\label{ssec:allegro}
\subsubsection{Setting of hyperparemters, $N_{\rm layer}$ and $l_{\rm max}$}
\label{ssec:hyper}
The Allegro model~\cite{Musaelian_2023} can be constructed using NequIP~\cite{Batzner_2022} which introduces an E(3) symmetry equivariant neural network (NN) coded using PyTorch~\cite{Paszke_2019}. 
However, unlike NequIP, Allegro questions the need for the message passing scheme that incorporates atomic interactions beyond the cutoff radius $R_{\rm c}$ and instead adopts the conventional method of restricting the interacting atoms to within $R_{\rm c}$ from the central atom.
The atomic potential energy of Allegro $\varepsilon^{\rm Allegro}_i$ is calculated as the sum of the pairwise energies $\varepsilon^{\rm Allegro}_{ij}$ defined between two atoms within $R_{\rm c}$:
\begin{eqnarray}
\varepsilon^{\rm Allegro}_i = \sum_{j} \varepsilon^{\rm Allegro}_{ij}
\label{eq:aenergy},
\end{eqnarray}
and the total potential energy is defined by the sum of $\varepsilon^{\rm Allegro}_i$ over all atoms in the system,
\begin{eqnarray}
E^{\rm Allegro} = \sum^{N_{\rm atom}}_{i} \varepsilon^{\rm Allegro}_i
\label{eq:energy}.
\end{eqnarray}
This definition of $E^{\rm Allegro}$ provides Allegro the advantage of parallel computation with space partitioning using the well-known Verlet neighbor list and the cell-index method required for large-scale MD simulations.

The motivation for developing Allegro was also to overcome the difficulties of ACE~\cite{Drautz_2019}.
ACE can construct descriptors that incorporate BO expansion up to higher orders at a feasible computational cost.
While most conventional descriptors are composed of up to three-body order of radial and angular parts, such as the Behler-Parrinello symmetry functions~\cite{Behler_2007}, ACE descriptors, which incorporates four-body order or more, have been employed~\cite{Araki_2023}.
Furthermore, ACE also efficiently generates descriptors with higher-order rotational symmetry characterized by irreducible representation $l$, which paved the way for MLIP construction with descriptors incorporating rotational equivariance~\cite{Batatia_2023}.
However, a drawback is that the number of descriptors increases exponentially as the number of atomic species increases.
In addition, the number of different descriptors, characterized by BO or $l$, to generate is determined by trial and error.
ACE descriptors are often combined nonlinearly to improve accuracy~\cite{Drautz_2019}, and the manner in which they are combined requires trial and error.

Allegro employs a GNN to distinguish atomic species by utilizing embedded vectors, which significantly reduce the computational cost resulting from an increase in atomic species.
Furthermore, the descriptor generation process is combined with the layer structure of GNN to efficiently construct nonlinear descriptors. 
The number of convolution layers in GNN, which is controlled by hyperparameter $N_{\rm layer}$, is devised to correspond to the order of BO expansion. 
As the tensor computation of the spherical harmonics is performed in each convolution layer of the GNN, the BO order of the descriptor increases by one.
For example, Allegro with $N_{\rm layer}$ = 2 produces a nonlinear combination of descriptors of up to four body orders. 
In other words, an Allegro model with $N_{\rm layer}$ has ($N_{\rm layer}$ + 2) body-order descriptors. 
In addition, the diversity of descriptors with irreducible representation $l$ can be increased through the composition of spherical harmonics via Clebsch–Gordan coefficients during the tensor computation. 
The hyperparameter $l_{\rm max}$ controls the maximum irreducible representation $l$ of the descriptors produced after composition.
For $l_{\rm max}$ = 0, only rotation-invariant descriptors are produced, whereas, for $l_{\rm max}$ $\geq$ 1, rotation-equivariant descriptors are also produced, which can improve feature extraction for the atomic local environment.

Therefore, because $N_{\rm layer}$ and $l_{\rm max}$ play important roles in Allegro, this study focuses on these two hyperparameters and evaluates their dependence on the thermal conductivity.
Six models are created using combinations of $N_{\rm layer}$ = \{1,2\} and $l_{\rm max}$ = \{0,1,2\}, that is, ($l_{\rm max}$,$N_{\rm layer}$) = (0,1), (0,2), (1,1), (1,2), (2,1), and (2,2) models.

Other hyperparameters related to the descriptors and architecture of GNN are the same for all Allegro models.
The details are explained in Section 2 of SM.

\subsubsection{Training of Allegro}
\label{ssec:training}
The Adam optimizer~\cite{Kingma_2017}, which is a stochastic gradient descent method, was adopted to train all the Allegro models.
The default parameters of Adam were set to $\beta_{1}$ = 0.9, $\beta_{2}$ = 0.999, and $\epsilon$ = 10$^{-8}$ without weight decay. 
The total number of epochs was 1,000.
All the models were trained with float64 precision by minimizing the following cost function $C$, which comprises three loss functions associated with the total potential energy (first term), atomic force (second term), and total virial (third term):
\begin{eqnarray}
C &=& \frac{p_E}{2}\frac{1}{B}\sum^{B}_{I} \left(  \frac{   E^{\rm Allegro}_I - E^{\rm FPMD}_I  }{N_{{\rm atom}}} \right)^2 \nonumber \\
&+& \frac{p_F}{2}\frac{1}{B}\sum^{B}_{I} \frac{1}{3N_{{\rm atom}}} \sum^{N_{{\rm atom}}}_{i}\ 
    \left({\bf{F}}^{\rm Allegro}_{I,i} - {\bf{F}}^{\rm FPMD}_{I,i} \right)^2 \nonumber \\
&+& \frac{p_W}{2}\frac{1}{B}\sum^{B}_{I} \frac{1}{6} \sum^{6}_{j}\ 
    \left( {W}^{\rm Allegro}_{I,j} - {W}^{\rm FPMD}_{I,j} \right)^2
\label{eq:cost},
\end{eqnarray}
where $B$ denotes the batch size used for the Adam optimizer and was set to $B$ = 5 in this study.
As these terms differ in dimension and size, $p_E$, $p_F$, and $p_W$ were introduced as adjustable parameters;
$p_E$ = $p_F$ = 1.0 and $p_W$ = 0.0005 were used.
Note that the training for ${\rm W}^{\rm Allegro}_{I}$ is equivalent to that of the potential part of virial stress tensor calculated by dividing ${\rm W}^{\rm Allegro}_{I}$ by the volume $\varOmega$.
The training was performed on 80\% of the data mentioned in Section~\ref{ssec:bag2se} with the remaining 20\% of the data used for validation.
Hereafter, the former and latter are referred to as the Train and Validation datasets, respectively.
The training was performed using System C at the Institute for Solid State Physics, the University of Tokyo.
The training wall times for the six Allegro models described in Section~\ref{ssec:hyper} are shown in Fig. S2(a) in SM.

There is a concern that the physical quantities calculated from the Allegro models depend on the initial weight parameters.
Therefore, for statistical evaluation, five models with different initial weight parameters were constructed.

\subsection{Thermal conductivity calculation by HNEMD method}
\label{ssec:hnemd}

\subsubsection{Atomic virial tensors for Allegro}
\label{ssec:avir}
The heat flux ${\bf{J}}_Q$ in Eq.~(\ref{eq:hflux}) requires an atomic virial tensor ${\rm W}_i$. However, this is not present in the original Allegro model~\cite{Musaelian_2023}.
The derivation of an atomic virial tensor that is appropriate for the heat flux requires careful consideration of the nature of Allegro as a many-body potential.
The extended code~\cite{Yu_2023} proposed by Yu et al. is helpful for this purpose.  
They introduced relative coordinates $\{{\bf{r}}_{ij}$ ($\equiv$ ${\bf{r}}_{j}$ $-$ ${\bf{r}}_{i}$)\} into the computational graph of the atomic potential energy $\varepsilon^{\rm Allegro}_i$ in PyTorch. 
This allows the calculation of its derivatives $\frac{\partial \varepsilon^{\rm Allegro}_i}{\partial {\bf{r}}_{ij}}$ by automatic differentiation, and Yu et al. defined the following type of atomic virial tensor ${\rm W}^{\rm sym}_i$:
\begin{eqnarray}
{\rm W}^{\rm sym}_i = \sum_{j \ne i} \frac{1}{2} \left[ {\bf{r}}_{ij} \otimes \frac{\partial \epsilon^{\rm Allegro}_i}{\partial {\bf{r}}_{ji}}
+ \left( {\bf{r}}_{ij} \otimes \frac{\partial \epsilon^{\rm Allegro}_i}{\partial {\bf{r}}_{ji}} \right)^{\rm T}
\right]
\label{eq:symwi_yu}. 
\end{eqnarray}
This definition is based on the objective of finding an atomic virial tensor that is symmetric as well as the total virial tensor.
In practice, this can be expressed as follows:
\begin{eqnarray}
{\rm W}^{\rm sym}_i = \sum_{j \ne i} {\bf{r}}_{ij} \otimes \frac{\partial \epsilon^{\rm Allegro}_i}{\partial {\bf{r}}_{ji}}
\label{eq:symwi}. 
\end{eqnarray}
However, ${\rm W}^{\rm sym}_i$ would not be suitable for thermal conductivity calculations.
According to the work by Fan {\it{et al.}} on Tersoff potential~\cite{Fan_2015}, the rigorous heat flux {\bf{J}}$_Q$ for many-body potentials can be derived from the following equation~\cite{Helfand_1960}.
\begin{eqnarray}
{\bf{J}}_Q &=& \frac{1}{\varOmega}\frac{d}{dt}\sum^{N_{\rm atom}}_{i} \left(t_i + \varepsilon_i \right) {\bf{r}}_{i} \nonumber \\
&=& \frac{1}{\varOmega}\sum^{N_{\rm atom}}_{i} \left(t_i + \varepsilon_i \right) {\bf{v}}_{i}
+ \frac{1}{\varOmega}\sum^{N_{\rm atom}}_{i} {\bf{r}}_{i} \frac{d}{dt}\left(t_i + \varepsilon_i \right)
\label{eq:origin}.
\end{eqnarray}
The second term of Eq.~(\ref{eq:origin}), which coincides with the third term of Eq.~(\ref{eq:hflux}), yields a rigorous formula for the atomic virial tensor.
As the atomic potential energy of Allegro $\epsilon^{\rm Allegro}_i$ is only a function of relative coordinates ${\bf{r}}_{ij}$ within a sphere of radius $R_{\rm c}$
\begin{eqnarray}
\epsilon^{\rm Allegro}_i = \epsilon^{\rm Allegro}_i \left( \{ {\bf{r}}_{ij} \}_{j \ne i} \right)
\label{eq:aefun},
\end{eqnarray}
the second term of Eq.~(\ref{eq:origin}) can be expressed as~\cite{Fan_2015}:
\begin{eqnarray}
\frac{1}{\varOmega}\sum^{N_{\rm atom}}_{i} {\bf{r}}_{i} \frac{d}{dt}\left(t_i + \varepsilon^{\rm Allegro}_i \right) = \frac{1}{\varOmega}
\sum^{N_{\rm atom}}_i
\sum_{j \ne i} \left(
{\bf{r}}_{ij} \otimes \frac{\partial \varepsilon^{\rm Allegro}_j}{\partial {\bf{r}}_{ji}} \right){\bf{v}}_i
\label{eq:jpot45}.
\end{eqnarray}
Thus, the following atomic virial tensor ${\rm W}^{\rm asym}_i$ would be appropriate for the heat flux~\cite{Shimamura_2021},
\begin{eqnarray}
{\rm W}^{\rm asym}_i = \sum_{j \ne i} {\bf{r}}_{ij} \otimes \frac{\partial \epsilon^{\rm Allegro}_j}{\partial {\bf{r}}_{ji}}
\label{eq:asymwi},
\end{eqnarray}
which is an asymmetric tensor.
Although the mathematical properties of the two atomic virial tensors are different, their summation over $N_{\rm atom}$ corresponds to the same total virial tensor ${\rm W}^{\rm Allegro}$.
To demonstrate that the derivation of ${\rm W}^{\rm asym}_i$ is indispensable, the thermal conductivities obtained using these two atomic virial tensors are compared later.
The code of the Allegro model, extended to the output ${\rm W}^{\rm asym}_i$ and ${\rm W}^{\rm sym}_i$, is available in Ref.~\cite{git01}.

\subsubsection{HNEMD method}
\label{ssec:hnemd_kappa}
The homogeneous non-equilibrium MD (HNEMD) method~\cite{Ciccotti_1979,Evans_1982,Yoshiya_2004}, which is based on the GK formula, successfully reduces the computational cost by enabling the calculation of the thermal conductivity $\kappa$ in the $x$ direction using the time average of the heat flux ${\bf{J}}_Q$ instead of integrating the autocorrelation function in Eq.~(\ref{eq:kappa}):
\begin{eqnarray}
   \kappa = \frac{\varOmega}{ F_{\rm ext}T} \lim_{t \to \infty} \langle J_{Q,x}\rangle_t
\label{eq:pkappa},
\end{eqnarray}
where $F_{\rm ext}$ denotes the magnitude of the perturbation along the $x$ direction of the system.
An appropriate $F_{\rm ext}$ is determined from the linear response regime by evaluating the thermal conductivity as a function of $F_{\rm ext}$.
Based on the evaluation results (Fig. S3 in SM), we selected $F_{\rm ext}$ = 0.005 bohr$^{-1}$ in this study.

\subsubsection{\label{ssec:shc}HNEMD-SHC method}
The SHC method, originally developed for the NEMD method~\cite{Zhou_2015}, was recently extended to HNEMD by Fan {\it et al.}~\cite{Fan_2019}.
First, we define the cross-correlation function ${\bf{K}}(t)$ between atomic enthalpy ${\rm h}_i$ and atomic velocity ${\bf{v}}_i$:
\begin{eqnarray}
{\bf{K}}(t) &=& \frac{1}{\varOmega}\sum^{N_{\rm atom}}_{i} {\rm h}_i(0) {\bf{v}}_i(t)
\label{eq:shc_cc}.
\end{eqnarray}
${\rm h}_i$ is given by: 
\begin{eqnarray}
{\rm h}_i &=& (t_i
+ {\varepsilon}_i){\rm I}
+ {\rm W}_i
\label{eq:shc_ent},
\end{eqnarray}
where I denotes the identity matrix.
Note that Fan {\it et al.} constructed a correlation function using only ${\rm W}_i$~\cite{Fan_2019}, whereas we used ${\rm h}_i$ for the construction.
When the Fourier transform of ${\bf{K}}(t)$ is expressed as
\begin{eqnarray}
{\tilde{\bf{K}}}(\omega) &=& \int^{\infty}_{-\infty} e^{i\omega t} {\bf{K}}(t) dt 
\label{eq:shc_fourier},
\end{eqnarray}
${\bf{K}}(t)$ can also be written using the inverse Fourier transform as
\begin{eqnarray}
{\bf{K}}(t) &=& \frac{1}{2\pi}\int^{\infty}_{-\infty} e^{-i\omega t} {\tilde{\bf{K}}}(\omega) d\omega \nonumber \\
&=& \frac{1}{2\pi} 2\ {\rm Re} \left[ \int^{\infty}_{0} e^{-i\omega t} {\tilde{\bf{K}}}(\omega) d\omega \right]
\label{eq:shc_ifourier},
\end{eqnarray}
where Re[$c$] is the real part of the complex number $c$.
As ${\bf{K}}(t = 0)$ of Eq.~(\ref{eq:shc_cc}) corresponds to the heat flux formula in Eq.~(\ref{eq:hflux}), the thermal conductivity $\kappa$ can be expressed as follows via Eqs.~(\ref{eq:pkappa}) and~(\ref{eq:shc_ifourier}):
\begin{eqnarray}
\kappa &=& \frac{\varOmega}{F_{\rm ext}T}K_{x}(0) \nonumber \\
&=& \frac{\varOmega}{F_{\rm ext}T}\int^{\infty}_{0} \frac{d\omega}{2\pi}
2\ {\rm Re} \left[ \tilde{K}_{x}(\omega) \right] 
\nonumber \\
&=& \int^{\infty}_{0} \frac{d\omega}{2\pi}
\tilde{\kappa}(\omega)
\label{eq:shc_form},
\end{eqnarray}
where $\tilde{\kappa}(\omega) \equiv \frac{2\varOmega}{F_{\rm ext}T}{\rm Re} \left[ \tilde{K}_{x}(\omega) \right] $; $K_{x}(0)$ and $\tilde{K}_{x}(\omega)$ denote the $x$ direction components of ${\bf{K}}(0)$ and ${\tilde{\bf{K}}}(\omega)$, respectively.
Thus, using spectral thermal conductivity $\tilde{\kappa}(\omega)$, the total thermal conductivity $\kappa$ can be decomposed in frequency space $\omega$.
For convenience, the cumulative $\kappa^{\rm cum}(\omega)$ is defined as:
\begin{eqnarray}
\kappa^{\rm cum}(\omega) &=& 
\int^{\omega}_{0} \frac{d\omega'}{2\pi}
\tilde{\kappa}(\omega')
\label{eq:shc_cumu}.
\end{eqnarray}

\subsubsection{HNEMD simulations with Allegro models}
\label{ssec:hnemd_simu}
All HNEMD simulations were executed using the $NVT$ ensemble.
The values of the cell vectors of the MD cell were averaged over those of the FPMDs with the $NPT$ ensemble that were used as the training data.
For each HNEMD simulation, after relaxation for 100,000 steps, a 1,000,000-MD step simulation was performed to sample ${\bf{J}}_Q$ of Eq.~(\ref{eq:hflux}) and calculate {\bf{K}}(t) in Eq.~(\ref{eq:shc_cc}).
Due to its anisotropic structure, the thermal conductivity of $\beta$-Ag$_2$Se may depend on the direction.
The thermal conductivity along the $a$-axis of the unit cell (see Fig. S1 in SM) was calculated as an example.
The thermal conductivities were calculated using Eqs.~(\ref{eq:pkappa}), (\ref{eq:shc_form}), and (\ref{eq:shc_cumu}).
HNEMD simulations were performed using System C at the Institute for Solid State Physics, the University of Tokyo.
The wall times of the simulations for the six Allegro models described in section~\ref{ssec:hyper} are shown in Fig. S2(a) in SM.

\section{\label{sec:results}Results and Discussion}

\subsection{\label{ssec:accuracy}The accuracy of Allegro models}
\begin{figure}[h]
\begin{center}  \includegraphics[width=10cm]{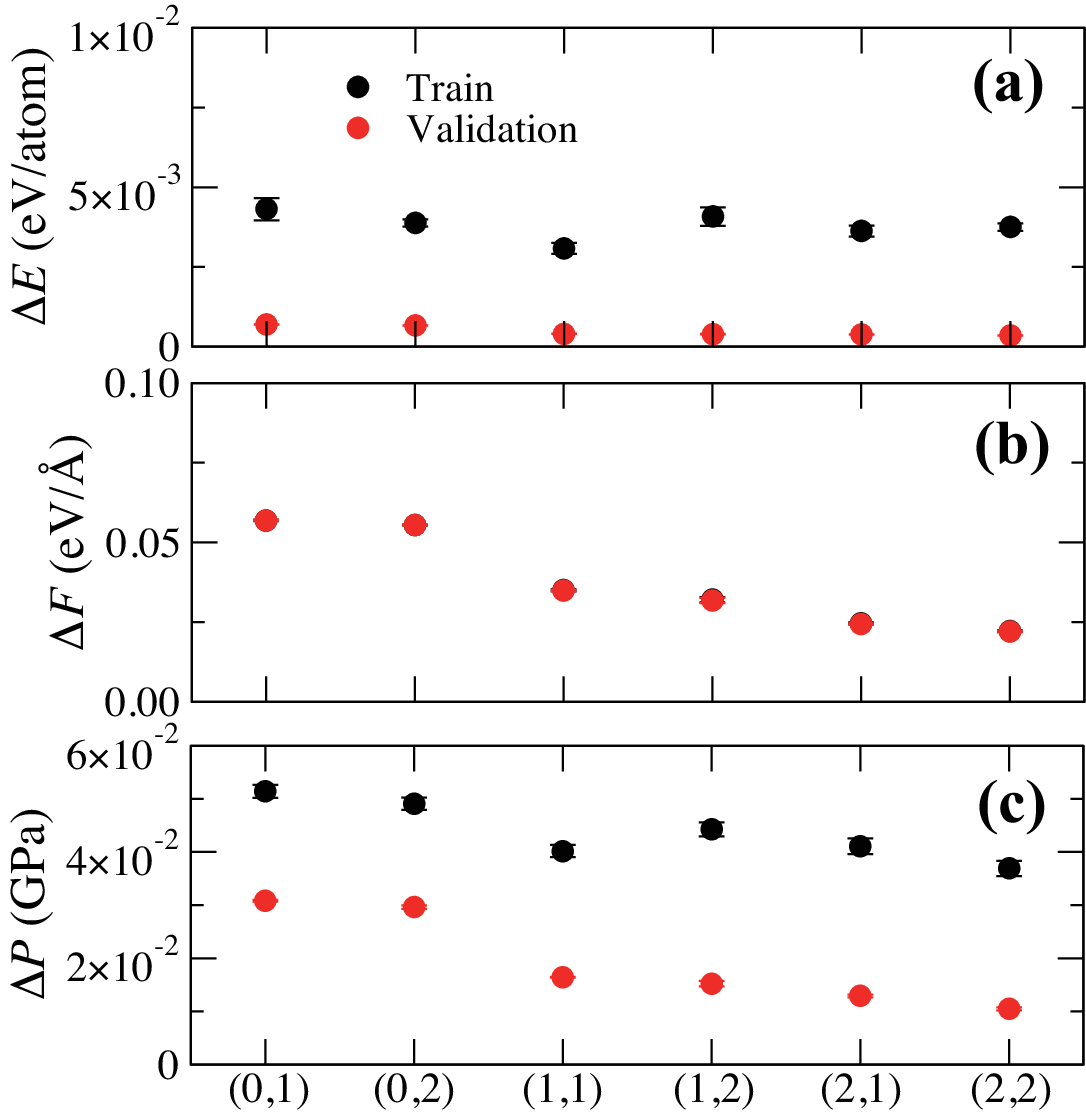}
  \caption{\label{Fig_RMSEs}Averaged root mean square errors and corresponding standard deviations (error bars) of (a) total potential energy ($\Delta E$), (b) atomic force ($\Delta F$), and (c) virial stress from the contribution of the potential energy ($\Delta P$) over the five Allegro models for each ($l_{\rm max}$,$N_{\rm layer}$) = (0,1), (0,2), (1,1), (1,2), (2,1), and (2,2). Black and red circles represent results obtained for the Train and Validation datasets, respectively.}
\end{center}
\end{figure}

First, we present the training results of the six Allegro models described in Section~\ref{ssec:hyper} for $\beta$-Ag$_2$Se., i.e., ($l_{\rm max}$,$N_{\rm layer}$) = (0,1), (0,2), (1,1), (1,2), (2,1), and (2,2) models. 
The root-mean-square errors (RMSEs) of the total potential energy ($\Delta E$), atomic force ($\Delta F$), and potential part of virial stress ($\Delta P$) are shown in Fig.~\ref{Fig_RMSEs}. 
All models showed low RMSEs of the orders of 10$^{-3}$ eV/atom for $\Delta E$, 10$^{-2}$ eV/\AA\ for $\Delta F$, and 10$^{-2}$ GPa for $\Delta P$, which are comparable to the RMSEs of silver chalcogenides in our previous studies using other types of MLIP~\cite{Shimamura_2021,Takeshita_2022,Shimamura_2024}.
However, nontrivial decreases in $\Delta F$ and $\Delta P$ are observed between the (0,2) and (1,1) models in Fig.~\ref{Fig_RMSEs}, which would cause discrepancies in some physical accuracy, such as atomic structure and stress, between these models. 
Hence, evaluating the accuracy of the atomic structure and stress through MD simulations is of interest.

\begin{figure}[h]
\begin{center}
  \includegraphics[width=10cm]{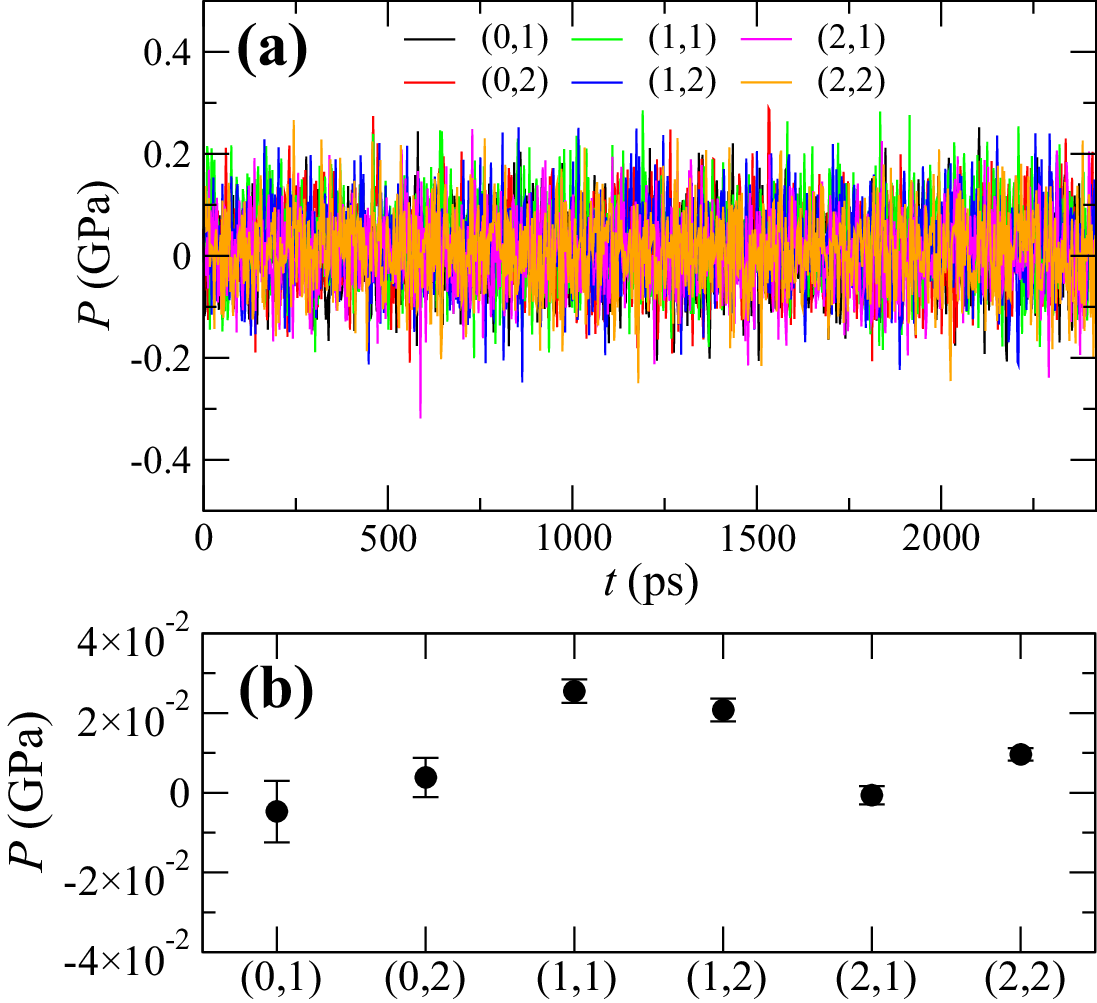}
  \caption{\label{Fig_Stress}
  (a) Stress profiles of the six Allegro models in the HNEMD simulation as a function of time. The results are shown for the Allegro models with the worst stress accuracy among the five models with different initial weight parameters. (b) The time-averaged stresses (the filled black circles) of the Allegro models with five different initial weight parameters for each ($l_{\rm max}$,$N_{\rm layer}$) = (0,1), (0,2), (1,1), (1,2), (2,1), and (2,2). Their standard deviations are represented by error bars.}
\end{center}
\end{figure}

Figure~\ref{Fig_Stress}(a) shows the time evolution of the stress through HNEMD simulations for each model. 
The stresses of all models oscillate around the training target of 0 GPa.
The time-averaged stresses for each Allegro model presented in Fig.~\ref{Fig_Stress}(b) are sufficiently close to 0 GPa and display no clear correlation with $\Delta P$ in Fig.~\ref{Fig_RMSEs}(c).
The differences observed in $\Delta P$ between the models are considered to have practically no effect on the stress in the MD simulations.
Incidentally, without virial training, the stress exhibits finite values, far from the target stress of 0 GPa ($\sim$6 GPa in the case shown in Fig. S4 in SM).
This indicates that virial training is essential for accurately reproducing the stress values in the Allegro model for $\beta$-Ag$_2$Se.
The accuracy of the virial stress would be important because it affects thermal conductivity via the atomic virial tensor~\cite{Shimamura_2021}.
 
\begin{figure}[h]
\begin{center}
  \includegraphics[width=12cm]{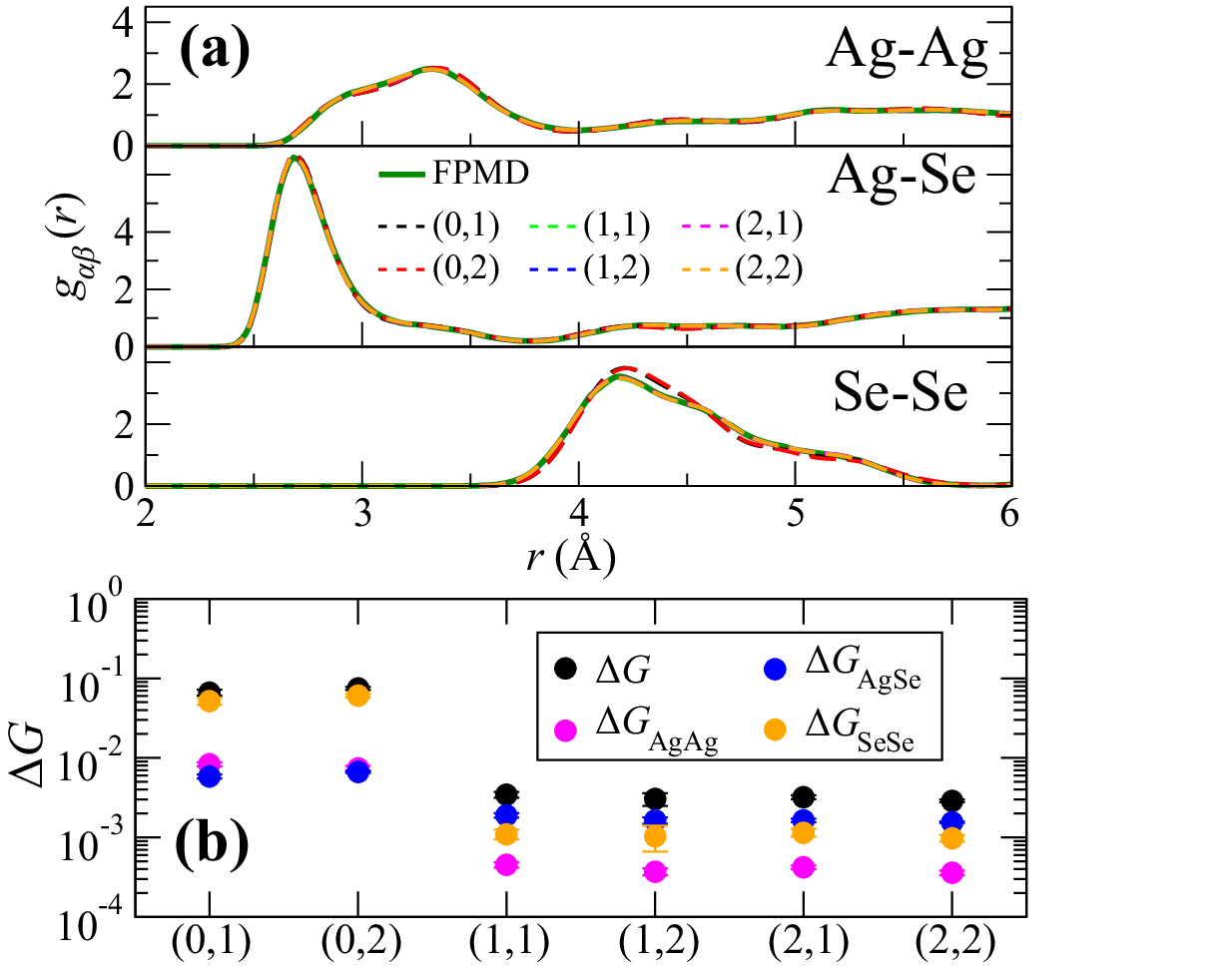}
  \caption{\label{Fig_gr}
  (a) Partial radial distribution functions $g_{\alpha \beta}$($r$) for Ag-Ag, Ag-Se, and Se-Se pairs obtained from FPMD (solid lines) and MD simulations using six Allegro models (dashed lines). $g_{\alpha \beta}$($r$) from Allegro models are averaged over five models with different initial values of the weight parameters.
  (b) Averaged residuals $\Delta G$ and their components ($\Delta G_{\rm AgAg}$, $\Delta G_{\rm AgSe}$, and $\Delta G_{\rm SeSe}$) over the five models with different initial weight parameters. Their standard deviations are represented by error bars. $\Delta G$ is defined in Eq.~(\ref{eq:DG}).}
\end{center}
\end{figure}

The reproducibility of the atomic structure with FPMD data was evaluated using partial radial distribution functions $g_{\alpha \beta}$($r$) for the Ag-Ag, Ag-Se, and Se-Se pairs.
Figure~\ref{Fig_gr}(a) shows $g_{\alpha \beta}$($r$) from FPMD and MD simulations using the Allegro models. 
We also quantified the correctness of $g_{\alpha \beta}$($r$) using the following $\Delta G$ from our previous studies~\cite{Irie_2022,Fukushima_2023}:
\begin{eqnarray}
\Delta G&=&\int_{0}^{R}
\left[ 
\left(\Delta g_{\rm AgAg}(r)\right)^2
+\left(\Delta g_{\rm AgSe}(r)\right)^2
+\left(\Delta g_{\rm SeSe}(r)\right)^2
\right]dr \nonumber \\
&=& \Delta G_{\rm AgAg}
+
\Delta G_{\rm AgSe}
+
\Delta G_{\rm SeSe}
\label{eq:DG},
\end{eqnarray}
where $\Delta g_{\alpha \beta}$($r$) denotes the difference between $g_{\alpha \beta}$($r$) obtained from the FPMD and MD simulations using the Allegro models.
$\Delta G_{\alpha \beta}$ is calculated by integrating $(\Delta g_{\alpha \beta}$($r$))$^2$ from 0 to $R$ = 6.0 \AA.
The values of $\Delta G$ and $\Delta G_{\alpha \beta}$ are presented in Fig.~\ref{Fig_gr}(b).
$\Delta G$ values of the (0,1) and (0,2) models are more than one order of magnitude larger than those of the other four models with $l_{\rm max}$ = 1 and 2.
This is because of the obvious deviation in $g_{\rm SeSe}$($r$) of the (0,1) and (0,2) models from that of FPMD in the region from the first peak position at 4.2 \AA\ to the subsequent 5.2 \AA\ in Fig.~\ref{Fig_gr}(a).
Since increasing $N_{\rm layer}$ from 1 to 2 does not change $g_{\rm SeSe}$($r$), increasing the BO of the descriptors did not lead to an essential improvement in reproducing the structure of Se. 
However, the remaining four models with $l_{\rm max}$ = 1 and 2 match to such an extent that no recognizable difference from FPMD is observed in Fig.~\ref{Fig_gr}(a). 
$\Delta G$ exhibits relatively lower values than the two models with $l_{\rm max}$ = 0. 
This improvement indicates that the dimensionality of the rotationally symmetric descriptors in the (0,1) and (0,2) models was insufficient.
Since the values of $\Delta G$ for the (1,1), (1,2), (2,1), and (2,2) models are almost the same, they reproduce $g_{\rm SeSe}$($r$) with similar accuracy.
Considering the computational cost (Fig. S2(a) in SM), we conclude that the (1,1) model is the most reasonable among the four models in terms of structural properties.

\subsection{\label{ssec:TC} 
Thermal conductivity}
\subsubsection{\label{sssec:wionTC}Effect of atomic virial tensor on thermal conductivity}
Figure~\ref{Fig_TC} shows the results of the thermal conductivities obtained from the HNEMD simulations using two types of atomic virial tensors ${\rm W}^{\rm sym}_i$ and ${\rm W}^{\rm asym}_i$ defined in Section~\ref{ssec:avir}.
As obtaining the reference value from FPMD is quite difficult because of the high computational cost, the thermal conductivities of the models were compared with the experimental value. 
The lattice thermal conductivity of $\beta$-Ag$_2$Se was experimentally estimated to be 0.3 $\pm$ 0.1 Wm$^{-1}$K$^{-1}$~\cite{Matsunaga_2021}, which is quite low despite the crystalline phase.
In terms of the results calculated using ${\rm W}^{\rm sym}_i$ in Eq.~(\ref{eq:symwi}), the four (0,1), (0,2), (1,1), and (1,2) models had almost the same thermal conductivities $\sim$0.45 Wm$^{-1}$K$^{-1}$, which were slightly larger than the experimental value.
However, the thermal conductivities of the (2,1) and (2,2) models further increased by $\sim$0.54 and $\sim$0.64 Wm$^{-1}$K$^{-1}$, respectively, clearly deviating from the experimental value.
By contrast, the thermal conductivities calculated using ${\rm W}^{\rm asym}_i$ in Eq.~(\ref{eq:asymwi}) are $\sim$0.35 Wm$^{-1}$K$^{-1}$ for the two models with $l_{\rm max}$ = 0, and $\sim$0.40 Wm$^{-1}$K$^{-1}$ for the four models with $l_{\rm max}$ = 1 and 2.
These values fall within the range of experimental error.
The fact that the former had a smaller thermal conductivity by $\sim$0.05 Wm$^{-1}$K$^{-1}$ would be attributed to the deviation of the structure from the FPMD simulation, however, the latter appeared to be converging, unlike those of ${\rm W}^{\rm sym}_i$.
From these observations, we conclude that our derivation of ${\rm W}^{\rm asym}_i$ would be essential for the thermal conductivity calculations using Allegro.

\subsubsection{\label{ssec:shcTC}Thermal conductivity analysis in frequency space}
\begin{figure}[h]
\begin{center}
  \includegraphics[width=10cm]{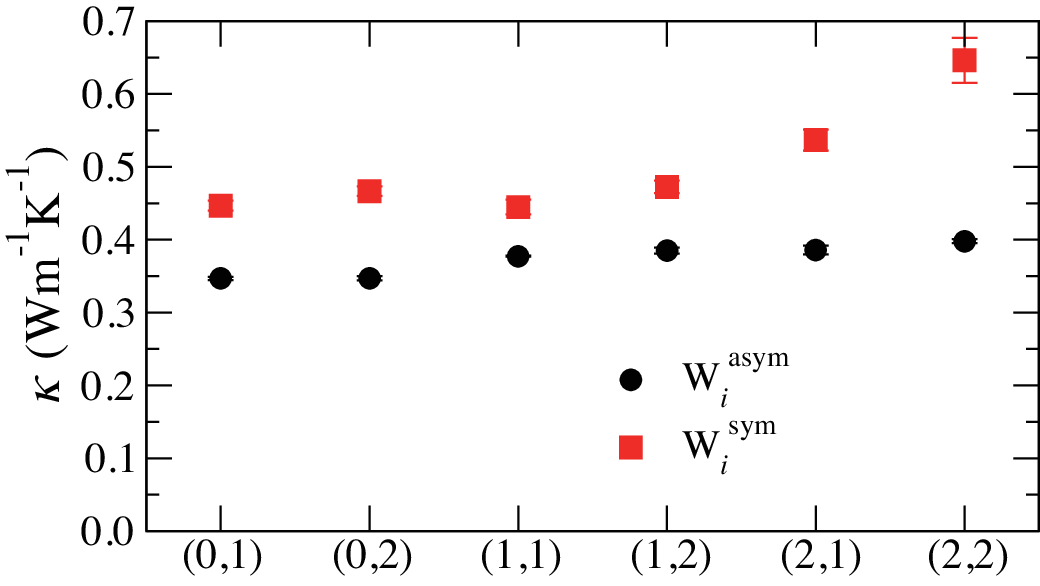}
  \caption{\label{Fig_TC}
  With two atomic virial tensors, ${\rm W}^{\rm asym}_i$ (black circles) and  ${\rm W}^{\rm sym}_i$ (red squares), the averaged thermal conductivities $\kappa$ are calculated from HNEMD simulations with $F_{\rm ext}$ = 0.005~bohr$^{-1}$ using six Allegro models. The averaged and corresponding standard deviations (error bars) are calculated from models with five different initial weight parameters.}
\end{center}
\end{figure}

\begin{figure}[h]
\begin{center}
  \includegraphics[width=10cm]{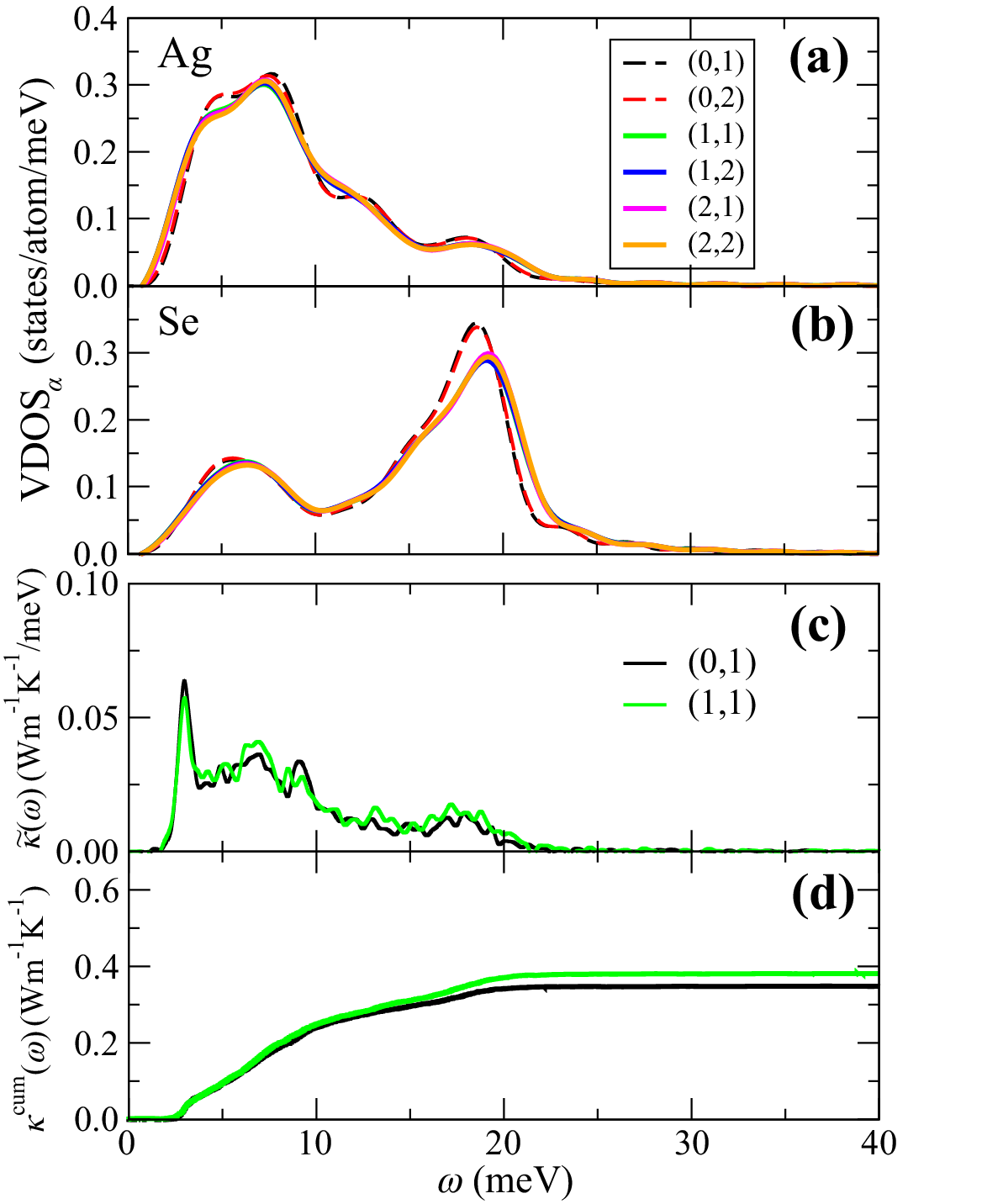}
  \caption{\label{Fig_freq}
  Elemental vibrational density of states (VDOSs) for (a) Ag and (b) Se  calculated from HNEMD simulations with the six Allegro models. (c) $\tilde{\kappa}(\omega)$ defined in Eq.~(\ref{eq:shc_form}) and (d) $\kappa^{\rm cum}(\omega)$ defined in Eq.~(\ref{eq:shc_cumu}) obtained by the HNEMD-SHC method using the (0,1) and (1,1) Allegro models. VDOSs, $\tilde{\kappa}(\omega)$, and $\kappa^{\rm cum}(\omega)$ are averaged over five Allegro models with different initial values of the weight parameters.}
\end{center}
\end{figure}

The increase in thermal conductivity by $\sim$0.05 Wm$^{-1}$K$^{-1}$ between the models with $l_{\rm max}$ = 0 and $l_{\rm max}$ = 1 and 2 (filled black circles in Fig.~\ref{Fig_TC}) would reflect the differences in the atomic structures, as shown in previous Section~\ref{ssec:accuracy}.
Using the HNEMD-SHC method~\cite{Fan_2019},
we can delve deeper into the cause of such small differences in thermal conductivity.
The differences in the atomic structures are reflected in the phonon distribution.
The vibrational density of states (VDOSs) has often been employed in many previous studies to elucidate the affected frequency regions~\cite{Loong_1992,Bryk_2020,Shenogin_2009,Varshney_2009,Minakov_2017,Zhou_2020}.
The partial VDOSs of Ag and Se (VDOS$_{\rm Ag}$ and VDOS$_{\rm Se}$) were calculated using the velocity correlation function~\cite{Loong_1992}, as shown in Figs.~\ref{Fig_freq}(a) and~\ref{Fig_freq}(b).
The profiles of VDOSs for the 
(0,1) and (0,2) models were similar.
However, the VDOSs for the (1,1), (1,2), (2,1), and (2,2) models were indistinguishable from one another.
Interestingly, a significant influence was observed not only on VDOS$_{\rm Se}$ but also on VDOS$_{\rm Ag}$, although the contributions of $\Delta G_{\rm AgAg}$ and $\Delta G_{\rm AgSe}$ to $\Delta G$ were quite small, as shown in Fig.~\ref{Fig_gr}(b).
The peak near 8 meV and shoulder near 4 meV in VDOS$_{\rm Ag}$ of the (0,1) model shift inward in the (1,1) model.
In addition, the peak near 12 meV in the (0,1) model had disappeared.
The peaks at approximately 5 and 18 meV for VDOS$_{\rm Se}$ in the (0,1) model shift slightly outward in the (1,1) model.
This may have caused the difference in total thermal conductivity; however, the contribution of each is unknown.

To this end, the HNEMD-SHC method is useful because the thermal conductivity can be decomposed in frequency space.
Figures~\ref{Fig_freq}(c) and~\ref{Fig_freq}(d) show the spectral thermal conductivity $\tilde{\kappa}(\omega)$ in Eq.~(\ref{eq:shc_form}) and its cumulative $\kappa^{\rm cum}(\omega)$ in Eq.~(\ref{eq:shc_cumu}), respectively, for the (0,1) and (1,1) models.
We found that $\tilde{\kappa}(\omega)$ from 15 to 20 meV in the (1,1) model was larger than that in the (0,1) model, which mainly contributes to the deviation in $\kappa$ 
$\sim$0.05 Wm$^{-1}$K$^{-1}$ in Fig.~\ref{Fig_TC}.
This can be understood more easily by $\kappa^{\rm cum}(\omega)$. 
Both models showed similar profiles from 0 to 15 meV.
After 15 meV, a difference emerged, and the values converged with a difference of $\sim$ 0.05 Wm$^{-1}$K$^{-1}$ after 20 meV.
The region from 15 to 20 meV corresponds to the peak shift of VDOS$_{\rm Se}$ in Fig.~\ref{Fig_freq}(b).

Considering these results, Allegro, which provides a high level description of atomic structures, is combined with the HNEMD-SHC method, which is expected to be even more useful for the thermal conductivity analysis of more complex systems, such as those with structural defects.

\section{\label{sec:conclusion}Conclusions}
The atomic virial tensor appropriate for the heat flux of the Allegro model was derived to calculate thermal conductivity using the HNEMD method based on the GK formula.
The derived atomic virial tensor works well for $\beta$-Ag$_2$Se as a test system, showing good agreement with the experimental value.
Even for Ag$_2$Se, with its extremely anharmonic atomic behavior, the relatively low descriptor levels of $l_{\rm max}$ = 1 and $N_{\rm layer}$ = 1 were sufficient.
However, the system used in this study was limited to a binary system with no defects.
By taking advantage of Allegro's ability to easily construct sophisticated descriptors, it would be possible to easily and accurately determine the thermal conductivities of a wide variety of materials with multicomponent complex structures that include defects.
In addition, using the recently developed HNEMD-SHC method, not only the total thermal conductivity but also the partial thermal conductivity of the frequency components can be obtained.
This would play an important role in clarifying the factors that generate thermal conductivity, as demonstrated in this study.
The effectiveness of the combination of Allegro with HNEMD and SHC methods will be further verified in the future.

\clearpage
\section*{CRediT authorship contribution statement}
{\bf{Kohei Shimamura:}} Methodology, Software, Formal analysis, Data curation, Writing - original draft, Visualization. 
{\bf{Shinnosuke Hattori:}} Conceptualization, Methodology, Software, Writing - review \& editing.
{\bf{Ken-ichi Nomura:}} Conceptualization, Methodology, Software, Writing - review \& editing.
{\bf{Akihide Koura:}} Conceptualization, Methodology, Writing - review \& editing, Supervision.
{\bf{Fuyuki Shimojo:}} Conceptualization, Methodology, Software, Resources, Writing - review \& editing, Supervision.

\section*{Declaration of Competing Interest}
The authors declare that they have no known competing financial interests or personal relationships that could have appeared to influence the work reported in this paper.

\section*{Acknowledgments}
This study was supported by MEXT/JSPS KAKENHI Grant Numbers Nos. 22K03454 and 21H01766, and JST CREST Grant Number JPMJCR18I2, Japan. The authors thank the Supercomputer Center, the Institute for Solid State Physics, University of Tokyo for the use of the facilities.

\section*{Data Availability}
An Allegro model is available at Ref.~\cite{git01} that allows the output of atomic virial tensors in ${\rm W}^{\rm asym}_i$ and ${\rm W}^{\rm sym}_i$.

\clearpage
\bibliographystyle{elsarticle-num-names}
%\bibliography{Manuscript}

\clearpage
\section*{Figures}

%\end{document}

%% file: Supplementary.tex
%\documentclass[preprint,review,12pt]{revtex4}
%\usepackage{amsmath}
%\usepackage{color}
%\usepackage{graphicx}
%\usepackage{multirow}
%\usepackage{array}
%\renewcommand{\figurename}{Figure S}
%\renewcommand{\tablename}{TABLE S}
%\newcolumntype{C}[1]{>{\hfil}m{#1}<{\hfil}}

%\makeatletter
%\renewcommand{\theequation}{
%    \thesection.\arabic{equation}}
%    \@addtoreset{equation}{section}
%\makeatother

%\begin{document}
%\title{Supplementary Material for ``Thermal Conductivity Calculation using Homogeneous Non-equilibrium Molecular Dynamics Simulation 
%with Allegro''}

%\author{Kohei Shimamura}
%\email{shimamura@kumamoto-u.ac.jp.}
%\affiliation{Department of Physics, Kumamoto University}

%\author{Shinnosuke Hattori}
%\affiliation{Advanced Research Laboratory, Technology Infrastructure Center, Technology Platform, Sony Group Corporation}

%\author{Ken-ichi Nomura}
%\affiliation{Collaboratory for Advanced Computing and Simulations, University of Southern California}

%\author{Akihide Koura}
%\affiliation{Department of Physics, Kumamoto University}

%\author{Fuyuki Shimojo}
%\affiliation{Department of Physics, Kumamoto University}

%\maketitle

%\setlength{\baselineskip}{24pt}

{\Large{Supplementary Material for ``Thermal Conductivity Calculation using Homogeneous Non-equilibrium Molecular Dynamics Simulation 
with Allegro''}}
\setcounter{section}{0}
\setcounter{figure}{0}
\setcounter{equation}{0}
\renewcommand{\figurename}{Figure S}
\renewcommand{\tablename}{TABLE S}

\clearpage
\section{Atomic configuration of $\beta$-Ag$_2$Se}
\begin{figure}[h]
\begin{center}
  \includegraphics[width=7cm]{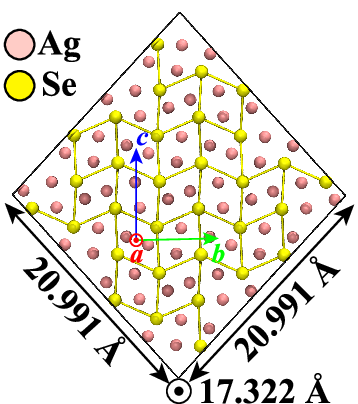}
  \caption{\label{FigS1}
  Atomic configurations of the systems consisting of 256 Ag (pink) and 128 Se (yellow) atoms for $\beta$-Ag$_2$Se. The arrows $a$, $b$, and $c$ represent orthorhombic unit cell vectors.}
\end{center}
\end{figure}

\clearpage
\section{Other hyperparameters related with descriptors and architecture of GNN}
We employed the cutoff radius $R_{\rm c}$ = 7 {\AA}.
A radial basis of eight trainable Bessel functions for the basis encoding with the polynomial envelope function using $p$ = 6 is employed.
The parameter {\it parity} is set to  {\it o3\_full}.
The 2-body latent multi-layer perceptron (MLP) consists of three hidden layers of dimensions [32, 64, 128], using SiLU nonlinearities~\cite{Elfwing_2017_forSM}. 
The later latent MLPs consist of one hidden layer of dimension 128, also using SiLU nonlinearities.
The final edge energy MLP has one hidden layer of dimension 32 with no nonlinearity.

\clearpage
\section{Wall times of tranining and HNEMD simulation}
As described in section 2.2.1 of the main text, we constructed six Allegro models with different $N_{\rm layer}$ and $l_{\rm max}$, i.e, ($l_{\rm max}$,$N_{\rm layer}$) = (0,1), (0,2), (1,1), (1,2), (2,1), and (2,2) models.
For these models, training and HNEMD simulations were performed using System C at the Institute for Solid State Physic, the University of Tokyo.
A single node GPU (NVIDIA A100) was used for training. 
The HNEMD simulations, on the other hand, were performed on our MD code QXMD~\cite{Shimojo_2019_2} using 128 CPU cores.
The wall times for the training and HNEMD calculations for the six models are shown in Fig. S2(a).
The number of weight parameters for the six models is also shown in Fig. S2(b).
The wall time (training + HNEMD simulation) strongly correlate with the number of weights of Allegro models. 
\begin{figure}[h]
\begin{center}
  \includegraphics[width=10cm]{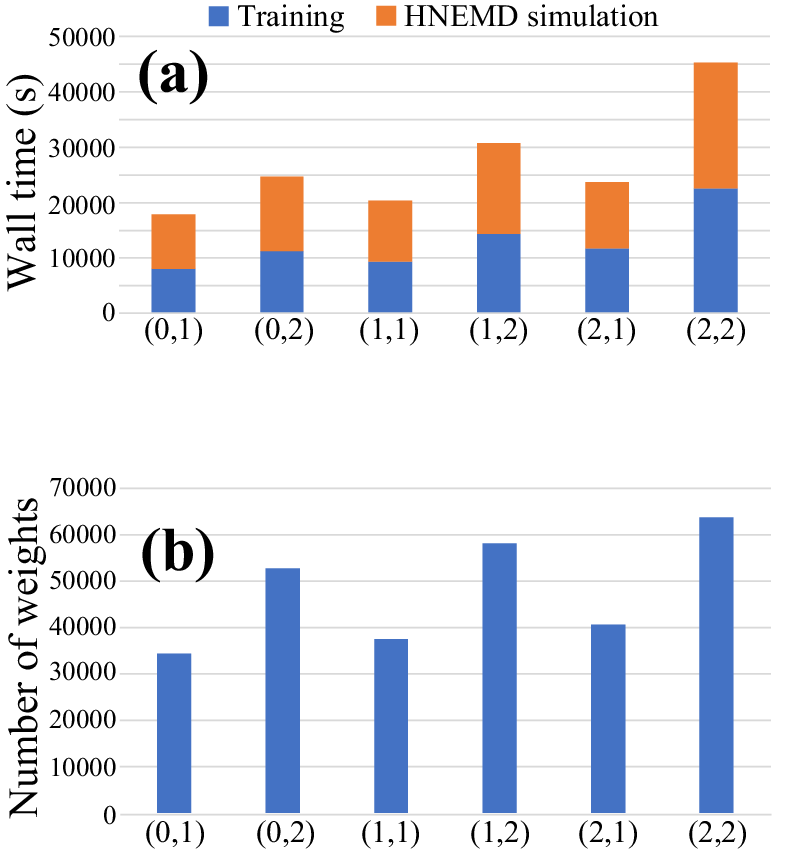}
  \caption{\label{FigS2}
  (a)Wall times of training and HNEMD simulation and (b)the numbers of weights parameters of six Allegro models, i.e, ($l_{\rm max}$,$N_{\rm layer}$) = (0,1), (0,2), (1,1), (1,2), (2,1), and (2,2) models. The wall times are averaged over five Allegro models with different initial values of the weight parameters.}
\end{center}
\end{figure}

\clearpage
\section{Determination of appropriate perturbation value}
Using the ($l_{\rm max}$,$N_{\rm layer}$) = (1,1) model as a representative, we determined the appropriate value of perturbation $F_{\rm ext}$ of HNEMD simulation, where the dependence of thermal conductivity $\kappa^{\rm Allegro}$ on $F_{\rm ext}$ ranged from 0.001 to 0.1~bohr$^{-1}$, as shown in Fig. S3.
The averaged thermal conductivities and corresponding standard deviations obtained among the five Allegro models with defferent initial weight parameters are represented by filled circles and error bars, respectively. 
We adopted $F_{\rm ext}$ = 0.005~bohr$^{-1}$, which has the lowest error bar.
\begin{figure}[h]
\begin{center}
  \includegraphics[width=10cm]{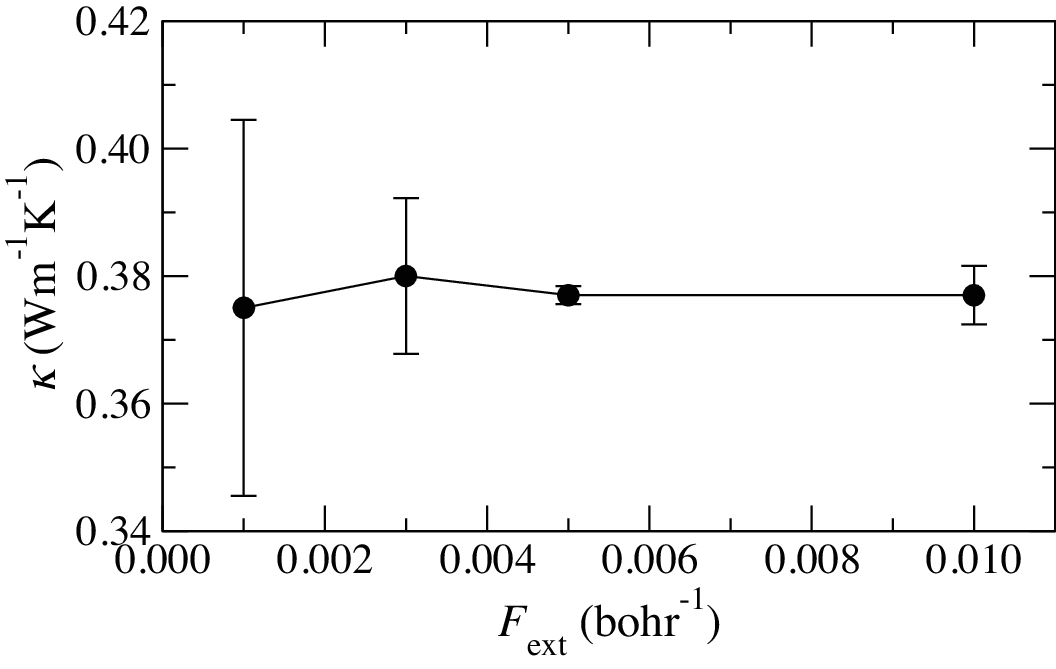}
  \caption{\label{FigS3}
  Averaged thermal conductivities calculated by HNEMD simulations using five Allegro models belonging to ($l_{\rm max}$,$N_{\rm layer}$) = (1,1) with different initial weight parameters as a function of the magnitude of the perturbation $F_{\rm ext}$. Error bars are illustrated using standard deviations of the thermal conductivities.}
\end{center}
\end{figure}

\clearpage
\section{Stress deviation seen in HNEMD simulation using Allegro model without virial training}
The stress behavior in the HNEMD simulation for the ($l_{\rm max}$,$N_{\rm layer}$) = (1,1) model without virial training (i.e., with $p_W$ = 0.0 in the cost function) is shown in Fig. S4.
The value clearly deviated from the target stress of training, 0 GPa.
The time-averaged stress of the Allegro models with five different initial weight parameters is 5.43 $\pm0.15$ 
 GPa.
\begin{figure}[h]
\begin{center}
  \includegraphics[width=10cm]{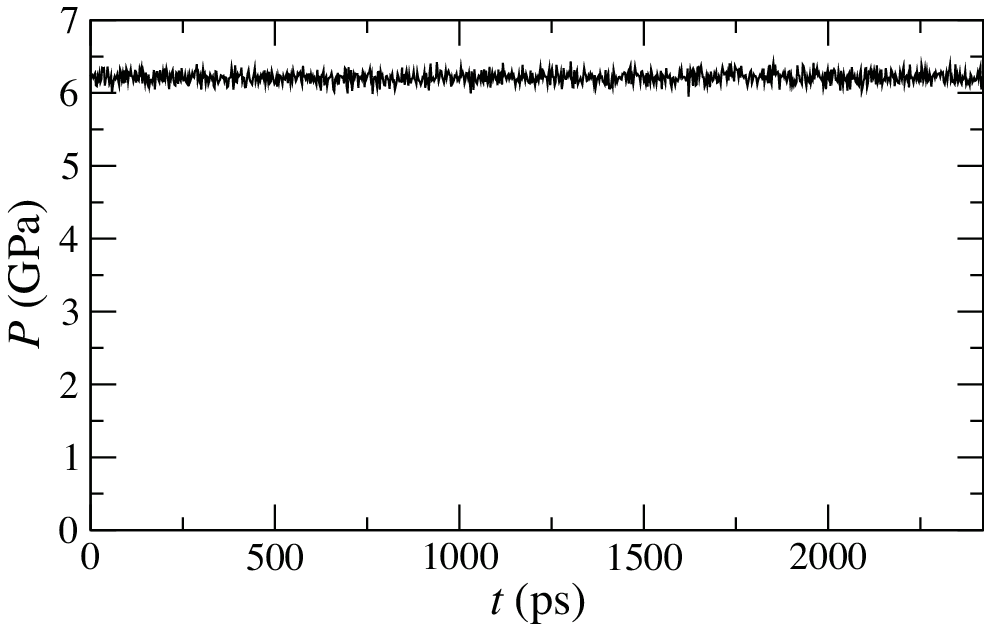}
  \caption{\label{FigS4}
  Stress profile of the (1,1) Allegro model without virial training in the HNEMD simulation as a function of time. The result is shown for the Allegro model with the smallest RMSE of virial stress from the contribution of the potential energy among the five models with different initial weight parameters.}
\end{center}
\end{figure}

\clearpage
%\section*{\large{References}}
%\bibliography{Manuscript}% Produces the bibliography via BibTeX.

%\end{document}